\documentclass[final]{eps}

\usepackage{graphicx}
\usepackage{times}
\usepackage{natbib}
\usepackage{amsmath,amssymb,amsfonts}

\volumenumber{58}
\publishyear{2006}
\frompage{\pageref{firstpage}}
\topage{\pageref{finalpage}}

\def\fuv{{\rm FUV}}
\def\tir{{\rm TIR}}
\def\pd{{\rm d}}
\def\vmatrix{{\mathbf{\Sigma}}}
\def\imatrix{{\mathbf{I}}}
\def\vx{\mathbf{x}}
\def\vpsi{\mathbf{\Psi^{-1}}}

\shorttitle{T. T. Takeuchi \lowercase{\textit{et al.}}: UV-IR Bivariate Luminosity Function of Galaxies}

\title{Far-Ultraviolet and Far-Infrared Bivariate Luminosity Function of Galaxies: Complex Relation between Stellar and Dust Emission}
\author{Tsutomu T. Takeuchi$^1$, 
Akane Sakurai$^1$, 
Fang-Ting, Yuan$^1$
V\'{e}ronique Buat$^2$,
and 
Denis Burgarella$^2$
}
\affiliation{$^1$Division of Particle and Astrophysical Science, Nagoya University, Furo-cho, Chikusa-ku, Nagoya 464-8602, Japan\\
             $^2$Laboratoire d'Astrophysique de Marseille, OAMP, Universit\'e Aix-Marseille, 
				CNRS, 38 rue Fr\'ed\'eric Joliot-Curie, 13388 Marseille cedex 13, France}
\received{xxxx xx, 2003}\revised{xxxx xx, 2003}\accepted{xxxx xx, 2003}
\abstract{
Far-ultraviolet (FUV) and far-infrared (FIR) luminosity functions (LFs) { of galaxies} 
show a strong evolution from $z = 0$ to $z = 1$, 
but the FIR LF evolves much stronger than the FUV one. 
The FUV is dominantly radiated from newly formed short-lived OB stars, 
while the FIR is emitted by dust grains heated by the FUV radiation field. 
It is known that dust is always associated with star formation activity. 
Thus, both FUV and FIR are tightly related to the star formation in galaxies, 
but in a very complicated manner. 
In order to disentangle the relation between FUV and FIR emissions, we estimate 
the UV-IR bivariate LF (BLF) of galaxies with {\sl GALEX} and {\sl AKARI} All-Sky Survey datasets. 
Recently we invented a new mathematical method to construct 
the BLF with given marginals and prescribed correlation coefficient. 
This method makes use of a tool from mathematical statistics, so called ``copula''. 
The copula enables us to construct a bivariate distribution function from 
given marginal distributions with prescribed correlation and/or dependence structure.
With this new formulation and FUV and FIR univariate LFs, we analyze various 
FUV and FIR data with {\sl GALEX}, {\sl Spitzer}, and {\sl AKARI} to estimate the UV-IR BLF. 
The obtained BLFs naturally explain the nonlinear complicated relation between 
FUV and FIR emission from star-forming galaxies. 
Though the faint-end of the BLF was not well constrained for high-$z$ samples,
the estimated linear correlation coefficient $\rho$ was found to be very high,
and is remarkably stable with redshifts (from 0.95 at $z = 0$ to 0.85 at $z = 1.0$).
This implies the evolution of the UV-IR BLF is mainly due to 
the different evolution of the univariate LFs, and may not be controlled by the dependence structure.
}

\begin{document}
\label{firstpage}
\maketitle
\copyrighttext{}

\section{Introduction}

Exploring the star formation history of galaxies is one of the most important 
topics in modern observational cosmology.
Especially, the ``true'' absolute value of the cosmic star formation rate
(hereafter SFR) has been of a central importance to understand the formation
and evolution of galaxies.

However, it has been a difficult task for a long time because of dust extinction.
Active star formation (SF) is always accompanied by dust production
through various dust grain formation processes related to the final stage of 
stellar evolution (e.g., Dwek and Scalo(1980); Dwek 1998; Nozawa et al.\ 2003; Takeuchi et al.\ 2005c; Asano et al.\ 2012).
Observationally, SFR of galaxies are, in principle, measured by the ultraviolet (UV) 
luminosity from massive stars because of their short lifetime ($\sim 10^6\;\mbox{yr}$) 
compared with the age of galaxies or the Universe.
However, since the UV photons are easily scattered and absorbed by dust grains,
SFR of galaxies is always inevitably affected by dust produced by 
their own SF activity.
Since the absorbed energy is re-emitted at wavelengths of far-infrared (FIR), 
we need observations both at UV and FIR to have an unbiased view
of their SF 
(e.g., Buat et al.\ 2005; Seibert et al.\ 2005; Cortese et al.\ 2006; Takeuchi et al.\ 2005a; 
Takeuchi et al.\ 2010; Haines et al.\ Bothwell et al.\ 2011; Hao et al.\ 2011).

After great effort of many researchers, the cosmic history of the SFR 
density is gradually converging at $0 < z < 1$. 
This ``latter half'' of the cosmic SFR is characterized by the rapid
decline of the total SFR, especially the decrease of the contribution
of dusty IR galaxies toward $z = 0$:
While at $z = 0$, the contribution of the SFR hidden by dust
is $50 \mbox{--} 60$~\%, it becomes $> 90$~\% at $z = 1$
(Takeuchi et al.\ 2005a).
This difference of decrease in SFR obtained from FUV and FIR has been
already recognized in IR studies (e.g., Takeuchi et al.\ 2001a,b).
Later works confirmed this gdusty era of the Universeh, and revealed that 
the dominance of the hidden SF continues even toward higher redshifts 
($z \sim 3$) (e.g., Murphy et al.\ 2011; Cucciati et al. 2012).

Then, a natural question arises: what does the different evolution at 
different wavelengths represent? 
{}To address this problem, it is very important to understand how we select 
sample galaxies and what we see in them. 
Each time we find some relation between different properties, we must understand 
clearly which is real (physical) and which is simply due to a selection effect. 
Some are detected at both bands, but some are detected only at one of the observed
wavelength and appear as upper limits at the other wavelength.
In previous studies it was often found that various claims were inconsistent with
each other, mainly because they did not construct a well-controled 
sample of FUV and FIR selected galaxies.
Recently, thanks to new large surveys, some attempts to explore the SFR distribution 
of galaxies in bivariate way have been made,  through the ``SFR function'' 
(e.g., Buat et al.\ 2007, 2009, Haines et al.\ 2011; Bothwell et al.\ 2011).
These works are based on the FUV and FIR LFs and their sum, but have not yet
address their dependence on each other.
{}To explore the bivariate properties of the SF in galaxies further, the proper tool 
is the UV-IR bivariate luminosity function (BLF). 

However, constructing a BLF from two-wavelength data is not a trivial task.
When we have a complete flux-limited\footnote{$^1$ Or any other observational 
condition.} multiwavelength dataset, we can estimate a univariate luminosity
function (LF) at each band, but what we want to know is {\it the dependence structure}
between luminosities at different bands.
Mathematically, this problem is rephrased as follows: 
can we (re)construct a multivariate probability 
density function (PDF) satisfying prescribed marginals?
Although there is an infinite number of degrees of freedom to choose the original PDF, 
if we can model the dependence between variables, we can construct such
a bivariate PDF.
A statistical tool for this problem is the so-called ``copula'' (see Sec.~\ref{sec:copula} for 
the definition).

A copula has been extensively used in financial engineering, for instance, but until recently
there were very few applications to astrophysical problems 
(e.g., Koen 2009; Benabed et al.\ 2009; Scherrer et al.\ 2010; Sato et al.\ 2010, 2011).
Takeuchi (2010) introduced the copula to the estimation problem of a BLF.
In this work, we apply the copula-based BLF estimation to the UV-IR 
two-wavelength dataset from $z=0$ to $z=1$, using data from {\sl IRAS}, 
{\sl AKARI}\footnote{$^2$ URL: http://www.ir.isas.ac.jp/ASTRO-F/index-e.html.}, 
{\sl Spitzer}\footnote{$^3$ URL: http://www.spitzer.caltech.edu/.}, and 
{\sl GALEX}\footnote{$^4$ URL: http://www.galex.caltech.edu/.}.

The layout of this paper is as follows.
We define a copula, especially the Gaussian copula, and formulate the copula-based BLF
in Sec.~\ref{sec:copula}. 
We then describe our UV and IR data in Sec.~\ref{sec:data}.
In Sec.~\ref{sec:discussion}, we first formulate the likelihood function for
the BLF estimation. 
Then we show the results, and discuss the possible interpretation of the
evolution of the UV-IR BLF.
Section~\ref{sec:conclusion} is devoted to summary.

Throughout the paper we will assume $\Omega_{\rm M0} = 0.3$, $\Omega_{\Lambda0} = 0.7$ 
and $ H_0 = 70 {\rm~ km~ s^{-1}~ Mpc^{-1}}$. 
The luminosities are defined as $\nu L_{\nu}$ and expressed  in solar units  assuming  
$L_{\odot} = 3.83 \times 10^{33} {\rm ~erg~ s^{-1}}$.

\section{The Bivariate Luminosity Function Based on the Copula}
\label{sec:copula}

\subsection{Copula: general definition}

First, we briefly introduce a copula.
A copula $C(u_1,u_2)$ is defined as follows:
\begin{eqnarray}\label{eq:copula}
  G(x_1, x_2) &=& C\left[ F_1 (x_1), F_2 (x_2) \right]
\end{eqnarray}
where $F_1(x_1)$ and $F_2(x_2)$ are two univariate marginal cumulative distribution 
functions (DFs) and $G(x_1, x_2)$ is a bivariate DF.
We note that all bivariate DFs have this form and we can safely apply this method to
any kind of bivariate DF estimation problem (Takeuchi 2010).
In various applications, we usually know the marginal DFs (or equivalently, PDFs) from 
the data. 
Then, the problem reduces to a statistical estimation of a set of parameters to
determine the shape of a copula $C(u_1,u_2)$.
In the form of the PDF, 
\begin{eqnarray}\label{eq:copula_density}
  g(x_1, x_2) &=& \frac{\partial^2 C\left[F_1 (x_1),F_2 (x_2)\right]}{\partial x_1 \partial x_2} 
    f_1 (x_1)  f_2 (x_2) \nonumber \\
    &\equiv& c\left[ F_1 (x_1), F_2 (x_2) \right] f_1(x_1) f_2 (x_2) ,
\end{eqnarray}
where $f_1 (x_1)$ and $f_2 (x_2)$ are the PDFs of $F_1(x_1)$ and $F_2(x_2)$, respectively.
On the second line, $c(u_1,u_2)$ is referred to as a differential copula.

\subsection{Gaussian copula}
\label{subsec:copula_gauss}

Since the choice of copula is literally unlimited, we have to introduce a guidance principle.
In many data analyses in physics, the most familiar measure of dependence 
might be the linear correlation coefficient $\rho$. 
Mathematically speaking, $\rho$ depends not only on the dependence of two 
variables but also the marginal distributions, which is not an ideal 
property as a dependence measure. 
Even so, a copula having an explicit dependence on $\rho$ would be convenient. 
In this work, we use a copula with this property, the Gaussian copula.

One of the natural candidate with $\rho$ may be a copula related to a bivariate Gaussian DF
(for other possibilities, see Takeuchi 2010).
The Gaussian copula has an explicit dependence on a linear correlation
coefficient by its construction.
Let
\begin{eqnarray}
  &&\psi_1 (x) = \frac{1}{\sqrt{2\pi}} \exp\left(-\frac{x^2}{2}\right), \label{eq:1gauss}\\
  &&\Psi_1 = \int_{-\infty}^{x} \Psi (x') \pd x' , \\
  &&\psi_2 (x_1, x_2; \rho) = \frac{1}{\sqrt{(2\pi)^2 \det \vmatrix}} 
    \exp\left(-\frac{1}{2} \vx^T \vmatrix^{-1} \vx \right) , \label{eq:2gauss_matrix} \nonumber \\
\end{eqnarray}
and
\begin{eqnarray}
  \Psi_2 (x_1, x_2; \rho) = \int_{-\infty}^{x_1} \int_{-\infty}^{x_2} \psi_i (x_1', x_2') \pd x_1' \pd x_2',
\end{eqnarray}
where $\vx \equiv (x_1, x_2)^T$, $\vmatrix$ is a covariance matrix 
\begin{eqnarray}\label{eq:vmatrix}
  \vmatrix \equiv  
    \begin{pmatrix}
      1 & \rho \\
      \rho & 1 
    \end{pmatrix} \;,
\end{eqnarray}
and superscript $T$ stands for the transpose of a matrix or vector.

Then, we define a Gaussian copula $C^{\rm G}(u_1, u_2; \rho)$ as
\begin{eqnarray}
  C^{\rm G} (u_1, u_2; \rho) = \Psi_2 \left[ \Psi_1^{-1} (u_1), \Psi_1^{-1} (u_2) ; \rho \right] \;.
\end{eqnarray}
The density of $C^{\rm G}$, $c^G$, is obtained as
\begin{eqnarray}
  &&c^{\rm G} (u_1, u_2; \rho) \nonumber \\
    &&\quad= 
    \frac{\partial^2 C^{\rm G} (u_1, u_2; \rho)}{\partial u_1 \partial u_2} \nonumber \\
    &&\quad= \frac{\partial^2 \Psi_2 \left[ \Psi_1^{-1} (u_1), \Psi_1^{-1} (u_2) ; \rho \right]}{
    \partial u_1 \partial u_2} \nonumber \\
    &&\quad= \frac{\psi_2 (x_1, x_2; \rho)}{\psi_1 (x_1) \psi_1 (x_2)} \nonumber \\
    &&\quad= \frac{1}{\sqrt{\det \vmatrix}}
    \exp\left\{-\frac{1}{2} \left[ \vpsi^T \left(\vmatrix^{-1} - \imatrix \right) \vpsi \right] \right\}, 
    \nonumber \\  
\end{eqnarray}
where $\vpsi \equiv \left[ \Psi^{-1}(u_1), \Psi^{-1}(u_2) \right]^T$ and $\imatrix$ stands for the identity
matrix.

\subsection{Construction of the UV-IR BLF}

In this work, we define the luminosity at a certain wavelength band by 
$L\equiv\nu L_\nu$ ($\nu$ is the corresponding frequency).
Then the LF is defined as the number density of galaxies whose
luminosity lies between a logarithmic interval
$[\log L, \log L + \pd\log L]$: 
\begin{eqnarray}
 \phi^{(1)} (L) \equiv \frac{\pd n}{\pd \log L} \;,
\end{eqnarray}
where we denote $\log x \equiv \log_{10} x$ and $\ln x \equiv \log_e x$.
For mathematical simplicity, we define the LF as being normalized, i.e., 
\begin{eqnarray}\label{eq:lfnorm}
 \int \phi^{(1)} (L) \pd\log L = 1\;.
\end{eqnarray}
Hence, this corresponds to a PDF.
We also define a cumulative LF as
\begin{eqnarray}
 \Phi^{(1)} (L) \equiv \int_{\log L_{\rm min}}^{\log L} \phi^{(1)} (L') \pd \log L' \;,
\end{eqnarray}
where $L_{\rm min}$ is the minimum luminosity of galaxies considered.
This corresponds to the DF.
If we denote the univariate LFs as $\phi^{(1)}_1(L_1)$ and $\phi^{(1)}_2(L_2)$, then
the BLF $\phi^{(2)}(L_1,L_2)$ is described by a differential copula $c(u_1, u_2)$ as
\begin{eqnarray}
  \phi^{(2)} (L_1, L_2) =
    c \left[\phi^{(1)}_1(L_1), \phi^{(1)}_2(L_2)\right] \;.
\end{eqnarray}
For the Gaussian copula, the BLF is obtained as
\begin{eqnarray}\label{eq:lf_gauss}
  &&\phi^{(2)}(L_1,L_2 ; \rho) \nonumber \\
    &&\quad =
    \frac{1}{\sqrt{\det \vmatrix}}
    \exp\left\{-\frac{1}{2} \left[ \vpsi^T \left(\vmatrix^{-1} - \imatrix \right) \vpsi \right] \right\} \nonumber \\
    &&\qquad \times \phi^{(1)}_1 (L_1) \phi^{(1)}_2 (L_2) \;,
\end{eqnarray}
where 
\begin{eqnarray}
  \vpsi = \left[ \Psi^{-1} \left( \Phi^{(1)}_1 (L_1) \right) , \; \Psi^{-1} \left( \Phi^{(1)}_2 (L_2) \right) \right]^T.
\end{eqnarray}

For the UV LF, we adopt the Schechter function (Schechter 1976).
\begin{eqnarray}\label{eq:schechter}
  \phi^{(1)}_1 (L) = (\ln 10)\; \phi_{*1} \left( \frac{L}{L_{*1}} \right)^{1-\alpha_1}
    \exp \left[-\left(\frac{L}{L_{*1}}\right)\right]. \nonumber \\
\end{eqnarray}
For the IR, we use the analytic form for the LF proposed by 
Saunders et al.\ (1990),
\begin{eqnarray}\label{eq:saunders}
  &&\phi^{(1)}_2 (L) \nonumber \\
    &&\quad = \phi_{*2} \left( \frac{L}{L_{*2}} \right)^{1-\alpha_2} \exp \left\{ -\frac{1}{2\sigma^2} 
    \left[ \log \left(1+\frac{L}{L_{*2}}\right)\right]^2\right\}.\nonumber \\
\end{eqnarray}
We use the re-normalized version of eqs.~(\ref{eq:saunders}) and (\ref{eq:schechter}) 
so that they can be regarded as PDFs, as mentioned above.

\subsection{Selection effects: another benefit of a copula BLF}

Another benefit of copula is that it is easy to incorporate observational selection effects
which always exist in any kind of astronomical data.
In a bi(multi)variate analysis, there are two categories of observational selection effects.
\begin{enumerate}
\item Truncation

We do not know if a source would exist below a detection limit.
\item Censoring

We know there is a source, but we have only an upper (sometimes lower) limit for a certain observable.
\end{enumerate}
As we mentioned above, we have to deal with both of these selection effects carefully to 
construct a  BLF from data at the same time.
It is terribly difficult to incorporate these effects by heuristic methods, particularly for
a nonparametric methods (Takeuchi 2012, in preparation).
In contrast, since we have an analytic form for the BLF, the treatment of upper limits 
is much more straightforward (Takeuchi 2010).
We show how it is treated in the likelihood function in Sec.~\ref{sec:discussion}.

\section{Data}
\label{sec:data}

\subsection{UV-IR data construction}

We have constructed a dataset of galaxies selected at FUV and FIR by
{\sl GALEX} and {\sl IRAS} for $z = 0$.
At higher redshifts, {\sl GALEX} and {\sl Spitzer} data are used 
for $z = 0.7$ and { EIS (ESO Imaging Survey)} and {\sl Spitzer} for $z=1.0$ in the Chandra 
Deep Field South (CDFS).
We explain the details of each sample in the following.

In the Local Universe, we used the sample compiled by Buat et al.\ (2007).
This sample was constructed based on {\sl IRAS} all-sky survey and {\sl GALEX} All-Sky Imaging 
Survey (AIS). 
This dataset consists of UV- and IR-selected samples.
The UV-selection was made by the {\sl GALEX} FUV ($\lambda = 1530$~\AA) band with 
$\text{FUV} < 17$~mag
(hereafter, all magnitudes are presented in AB mag).
This corresponds to the luminosity lower limit of $L_{\fuv} > 10^8 \; L_\odot$.
Redshift information was taken from HyperLEDA (Paturel et al.\ 2003) and NED.
The IR-selection is based on the {\sl IRAS} PSC$z$ (Saunders et al.\ 2000).
The detection limit of the PSC$z$ is $S_{60} > 0.6$~Jy.
Redshifts of all PSC$z$ galaxies were measured, and most of their redshifts are 
$z < 0.05$. 
All UV fluxes were remeasured with the package we have developed
(Iglesias-P\'{a}ramo et al.\ 2006) to avoid the shredding of galaxies.
Details of the sample construction are explained in Buat et al.\ (2007).
The number of the UV- and IR-selected samples are 606 and 644, respectively. 

We also constructed a new, much larger sample of IR-selected galaxies
by {\sl AKARI} FIS All-Sky Survey.
We started from the {\sl AKARI} FIS bright source catalog (BSC) v.$1$ from the {\sl AKARI} all sky survey
(Yamamura et al.\ 2010).
This sample is selected at {\it WIDE-S} band ($\lambda = 90\;\mu$m) of the {\sl AKARI} FIS 
(Kawada et al.\ 2007).
The detection limit is $S_{90} > 0.2$~Jy.
We first selected {\sl AKARI} sources in the SDSS footprints { to have the same solid angle with
a forthcoming corresponding UV-selected sample we are preparing with {\sl GALEX}-SDSS.\footnote{$^5$
This step is not a mandatory in this study, but we are planning to make an extension of this analysis 
with the UV-selected data, and this step will make the analysis
easier with respect to the treatment of survey areas when the UV-selected data will be ready.}
}
Then, to have a secure sample of galaxies with redshift data, we made a cross match of {\sl AKARI} 
sources with the Imperial {\sl IRAS}-FSC Redshift Catalogue (IIFSCz), a redshift catalog recently 
published (Wang and Rowan-Robinson 2009), with a search radius of 36~arcsec.
Since about 90~\% of galaxies in the IIFSCz have spectroscopic or photometric
redshifts at $S_{60} > 0.36$~Jy, the depth of the sample is defined by 
this matching.
This determines the redshift range of this dataset, to be approximately $z < 0.1$.
We measured the FUV and NUV flux densities with the same procedure as the {\sl IRAS}-{\sl GALEX}
sample.
The detection limits at FUV and NUV of this sample are $19.9$ mag and $20.8$ mag 
(Morrissey et al.\ 2007).
A corresponding UV-selected sample is under construction, hence
we have only IR-selected sample. 
The number of galaxies is 3,891.
For more information on this sample and properties of galaxies, see Sakurai et al.\ (2012).

At higher-$z$, our samples are selected in the CDFS.
{\sl GALEX} observed this field at FUV and NUV (2300~\AA) as a part of its deep imaging survey.
We restricted the field to the subfield observed by {\sl Spitzer}/MIPS as a part of the GOODS
key program (Elbaz et al.\ 2007) to have the corresponding IR data.
The area of the region is $0.068\;\text{deg}^2$.
Precise description of our high-$z$ samples are presented in Buat et al.\ (2009) and Burgarella et al.\ (2006).

At $z = 0.7$, the NUV-band corresponds to the { rest-frame FUV} of the sample galaxies.
We thus constructed the sample based on NUV-selection.
Redshifts were taken from the COMBO-17 survey (Wolf et al.\ 2004), and
we have defined the $z = 0.7$ sample as those with redshifts of $0.6 < z < 0.8$.
Data reduction and redshift association are explained in Burgarella et al.\ (2006).
We truncated the sample at NUV = 25.3~mag so that more than 90~\% of 
the {\sl GALEX} sources are identified in COMBO-17 with $R < 24$~mag.
We set the MIPS  $24\;\mu$m upper limit as 0.025~mJy for undetected sources.
For the IR-selected sample, we based on the GOODS {\sl Spitzer}/MIPS 24~$\mu$m sample 
and matched the {\sl GALEX} and COMBO-17 sources. 
The sizes of the UV- and IR-selected samples are 340 and 470, respectively.

In contrast to $z=0.7$, since NUV of {\sl GALEX} corresponds to 1155~\AA\ in
the { rest frame} of galaxies at $z=1.0$, we cannot use NUV as the primary selection
band as rest-frame FUV.
Instead, we selected galaxies in the $U$-band from the EIS survey 
(Arnouts et al.\ 2001) that covers the CDFS/GOODS field.
We then cross-matched the $U$-band sources with the COMBO-17 sample to
obtain redshifts.
For the $z = 1$ sample, the range of redshifts is $0.8 < z < 1.2$. 
We set the limit at $U = 24.3$~mag to avoid source confusion.
The IR flux densities were taken from {\sl Spitzer} MIPS $24\;\mu$m data, and
the same upper limit { as the $z = 0.7$ sample} is put for non-detections.
Again the IR-selected sample was constructed from the GOODS and matched 
the {\sl GALEX} and COMBO-17 sources, as the $z=0.7$ 
sample.
The sizes of the $z = 1.0$ UV and IR-selected samples are 319 and 1033, 
respectively.

{ The characteristics of these samples are summarized in Table~\ref{tab:sample}. }

\begin{table*}[tb]
\begin{center}
{ 
\caption{Sample description}\label{tab:sample}
\begin{tabular}{llll} \hline \hline 
Redshift & [range] & Primary selection & Number \\
\hline
$z = 0$ & $[0, 0.05]$ & UV ({\sl GALEX} FUV) & 606 \\
           &                & IR ({\sl IRAS} 60~$\mu$m) & 644 \\
           & $[0, 0.1]$  & IR ({\sl AKARI} {\sl WIDE-S}) & 3891 \\ \hline
$z = 0.7$ & $[0.6, 0.8]$ & UV ({\sl GALEX} NUV) & 340 \\
             &                  & IR ({\sl Spitzer} MIPS $24\;\mu$m) & 470 \\ \hline
$z = 1.0$ & $[0.8, 1.2]$ & UV (EIS $u$-band) & 319 \\
             &                  & IR ({\sl Spitzer} MIPS $24\;\mu$m) & 1033 \\
\hline
\end{tabular}
}
\end{center}
\end{table*}

\subsection{Far-UV and total IR luminosities}

We are interested in the SF activity of galaxies and its evolution.
Hence, luminosities of galaxies representative of SF activity would be
ideal.
As for the directly visible SF, obviously the UV emission is appropriate
for this purpose.
We define the FUV luminosity of galaxies $L_{\rm FUV}$ as $L_{\rm FUV}
\equiv \nu L_\nu\text{@FUV}$.
For $z=0$ galaxies, FUV corresponds to 1530~\AA.
At higher redshifts, as we explained, $L_{\rm FUV}$ is calculated from 
the NUV flux density at $z = 0.7$ and $U$-band flux density at $z = 1.0$, 
respectively.

In contrast, at IR, the luminosity related to the SF activity is the one integrated
over a whole range of IR wavelengths ($\lambda = 8 \mbox{--} 1000\;\mu$m), 
$L_{\rm TIR}$, { where the subscript TIR stands for the total IR}. 
For the Local {\sl IRAS}-{\sl GALEX} sample, it is quite straightforward to define $L_{\rm TIR}$ since
the {\sl IRAS} galaxies are selected at $60\;\mu$m.
We adopted a formula $L_{\tir} = 2.5 \nu L_\nu \text{@60~$\mu$m}$.
This rough approximation is, in fact, justified by spectral energy distributions (SEDs) 
of galaxies with {\sl ISO} $160\;\mu$m observations (Takeuchi et al.\ 2006).

Also, for the {\sl AKARI}-{\sl GALEX} sample, we have excellent flux density data at FIR.
We used the following TIR estimation formula from {\sl AKARI} two wide bands 
(Takeuchi et al.\ 2005b, 2010),
\begin{eqnarray}
L_{\rm TIR} &=& \Delta\nu({\mbox{\it{WIDE-S}}})L_{\nu}(90\;\mu {\rm m}) \nonumber \\
&&\qquad+ \Delta\nu({\mbox{\it{WIDE-L}}})L_{\nu}(140\;\mu {\rm m}) ,
\label{AKARI2band_luminosity}
\end{eqnarray}
where
\begin{equation}
\begin{split}
&\Delta\nu({\mbox{\it{WIDE-S}}}) = 1.47 \times 10^{12}\;[\rm H \rm z] \\
&\Delta\nu({\mbox{\it{WIDE-L}}}) = 0.831 \times 10^{12}\;[\rm H \rm z] . 
\label{AKARI3band_luminosity}
\end{split}
\end{equation}

However, since deep FIR data of higher-$z$ galaxies are not easily available to date,
we should rely on the data taken by {\sl {\sl Spitzer}} MIPS $24\;\mu$m.
we use conversion formulae from MIR luminosity to $L_{\rm TIR}$:
\begin{eqnarray}
  \log L_{\rm TIR} [\mbox{L}_\odot ] &=& 1.23 + 0.972 \log L_{15} [\mbox{L}_\odot ] \;,\\
  \log L_{\rm TIR} [\mbox{L}_\odot ] &=& 2.27 + 0.707 \log L_{12} [\mbox{L}_\odot ] \nonumber \\
  &&\qquad + 0.0140 \left( \log L_{12} [\mbox{L}_\odot ] \right)^2 , 
\end{eqnarray}
which are an updated version of the formulae proposed by Takeuchi et al.\ (2005b) and also used by
Buat et al.\ (2009).
Here $L_{15}$ and $L_{12}$ are luminosities $\nu L_\nu$@$15\mu$m and $12\mu$m,
i.e., $24/(1+z)$ at $z = 0.7$ and 1.0, respectively.
The estimated $L_{\tir}$ slightly depends on which kind of conversion formula is used,
but for our current purpose, it does not affect our conclusion and we do not discuss 
extensively here. 
Intercomparison of the MIR-TIR conversion formulae can be found in Buat et al.\ (2009).

This situation will be greatly improved with {\sl Herschel}\footnote{$^6$ URL: http://herschel.esac.esa.int/.} data.
We will leave this as our future work with {\sl Herschel} H-ATLAS (Eales et al.\ 2010) and 
GAMA (Galaxy And Mass Assembly: Driver et al.\ 2011)
data (Takeuchi et al. 2012, in preparation).

\begin{figure*}[t]
\centering\includegraphics[width=0.45\textwidth]{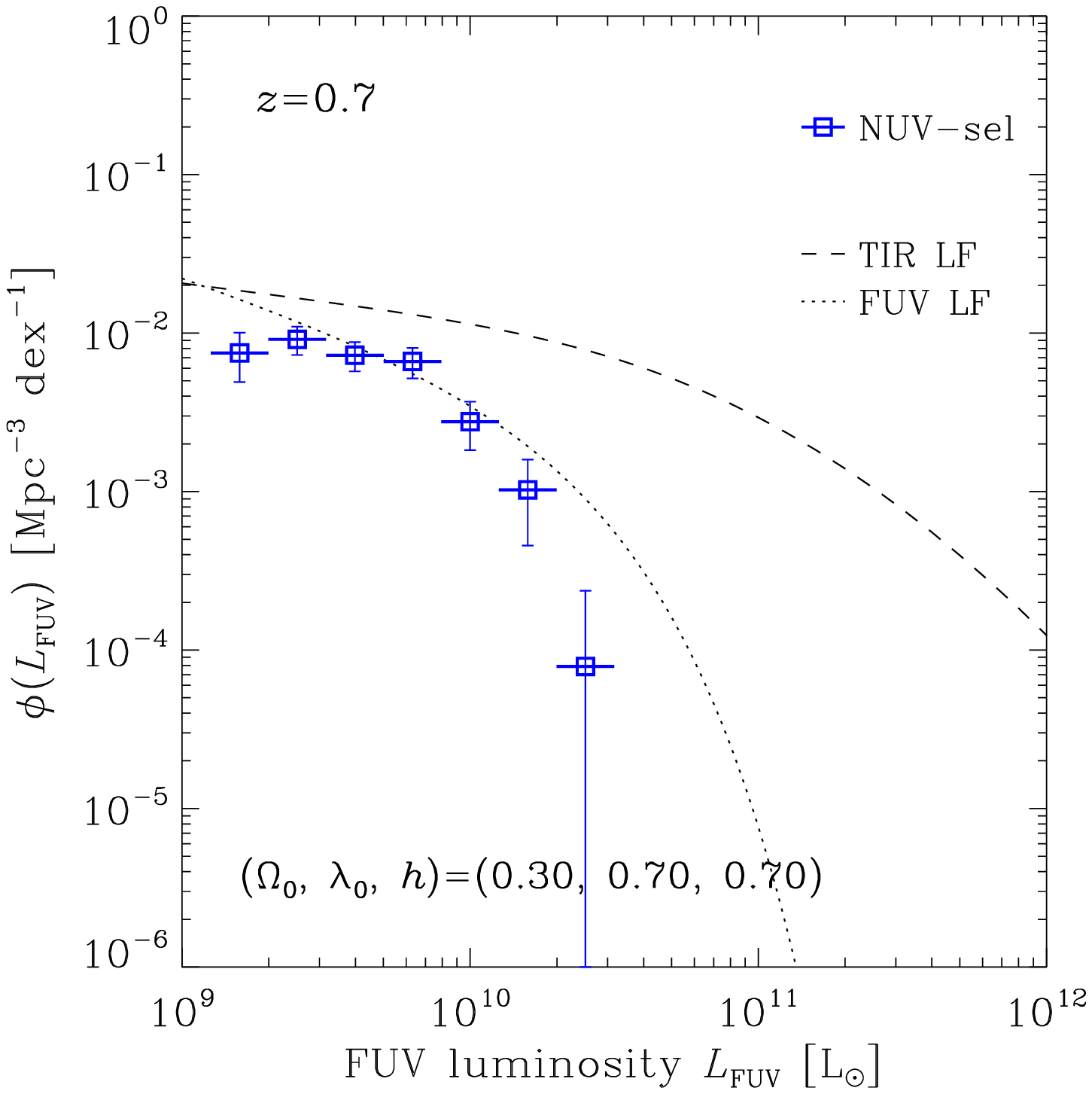}
\centering\includegraphics[width=0.45\textwidth]{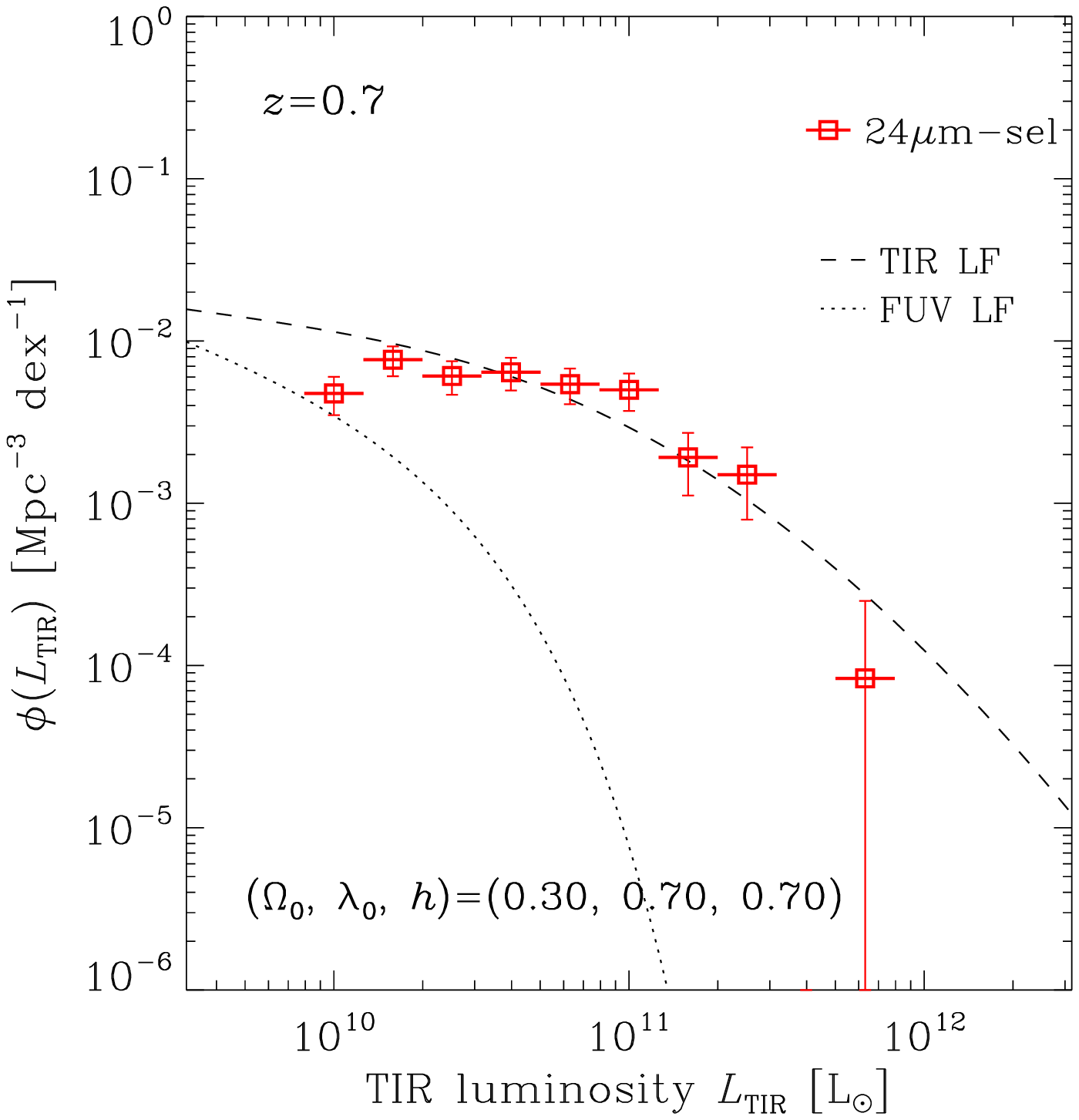}
\caption{The FUV and TIR LFs at $z = 0.7$ (left and right panel, respectively).
Open squares are a nonparametric LFs estimated from our CDFS multiwavelength data
with {\sl GALEX} NUV- and {\sl Spitzer} MIPS $24\;\mu$m-selections.
Dotted lines represent the FUV LFs at this redshift bin taken from Arnouts et al.\ (2005).
Dashed lines depict the TIR LFs derived from the evolutionary parameters at $z=0.7$
given by Le Floc'h et al.\ (2005).
Because of a well-known large density enhancement at this redshift, we renormalized the LF to
remove the effect of the overdensity.
}\label{fig:LF07} 
\end{figure*}
\begin{figure*}[t]
\centering\includegraphics[width=0.45\textwidth]{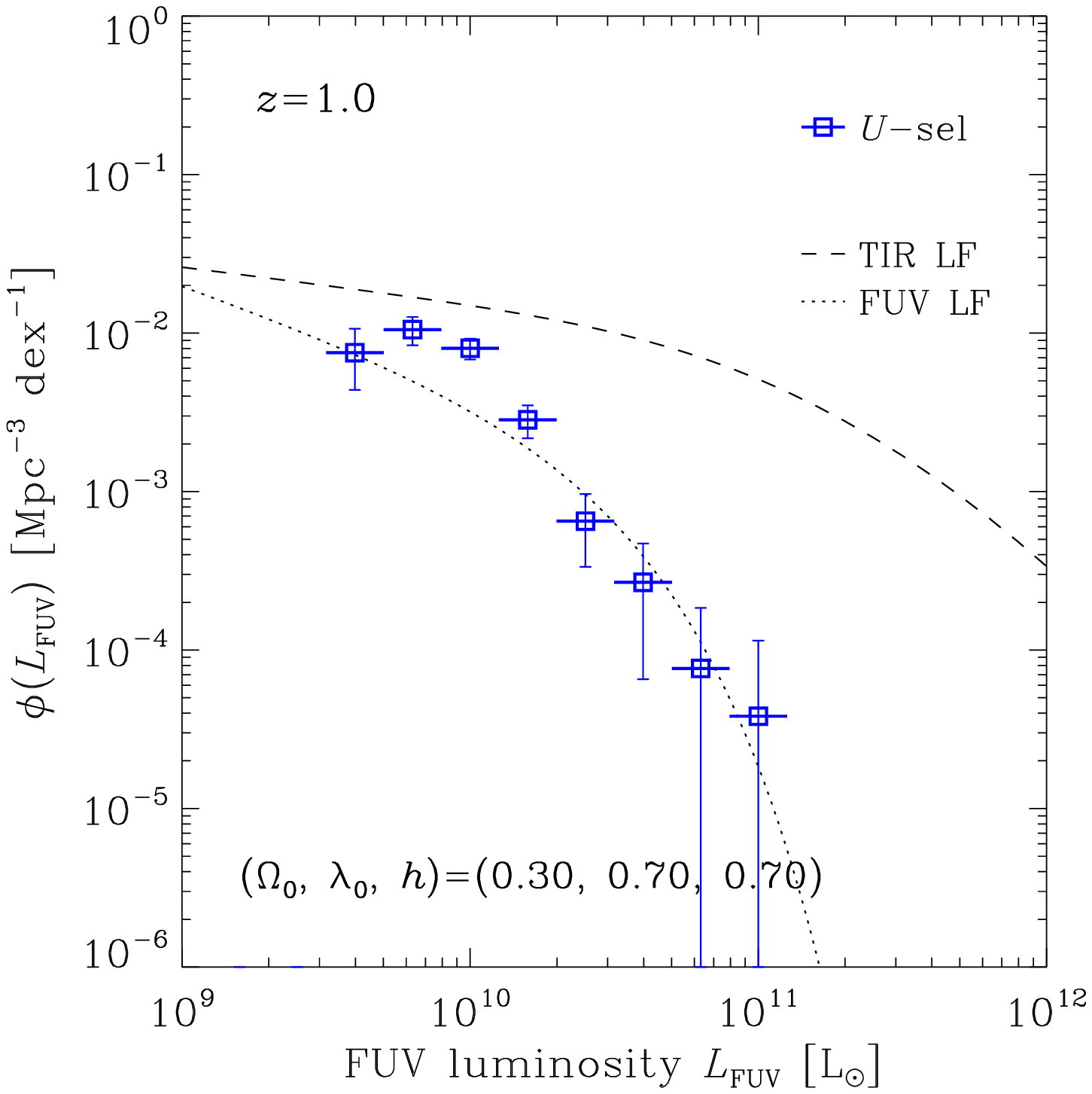}
\centering\includegraphics[width=0.45\textwidth]{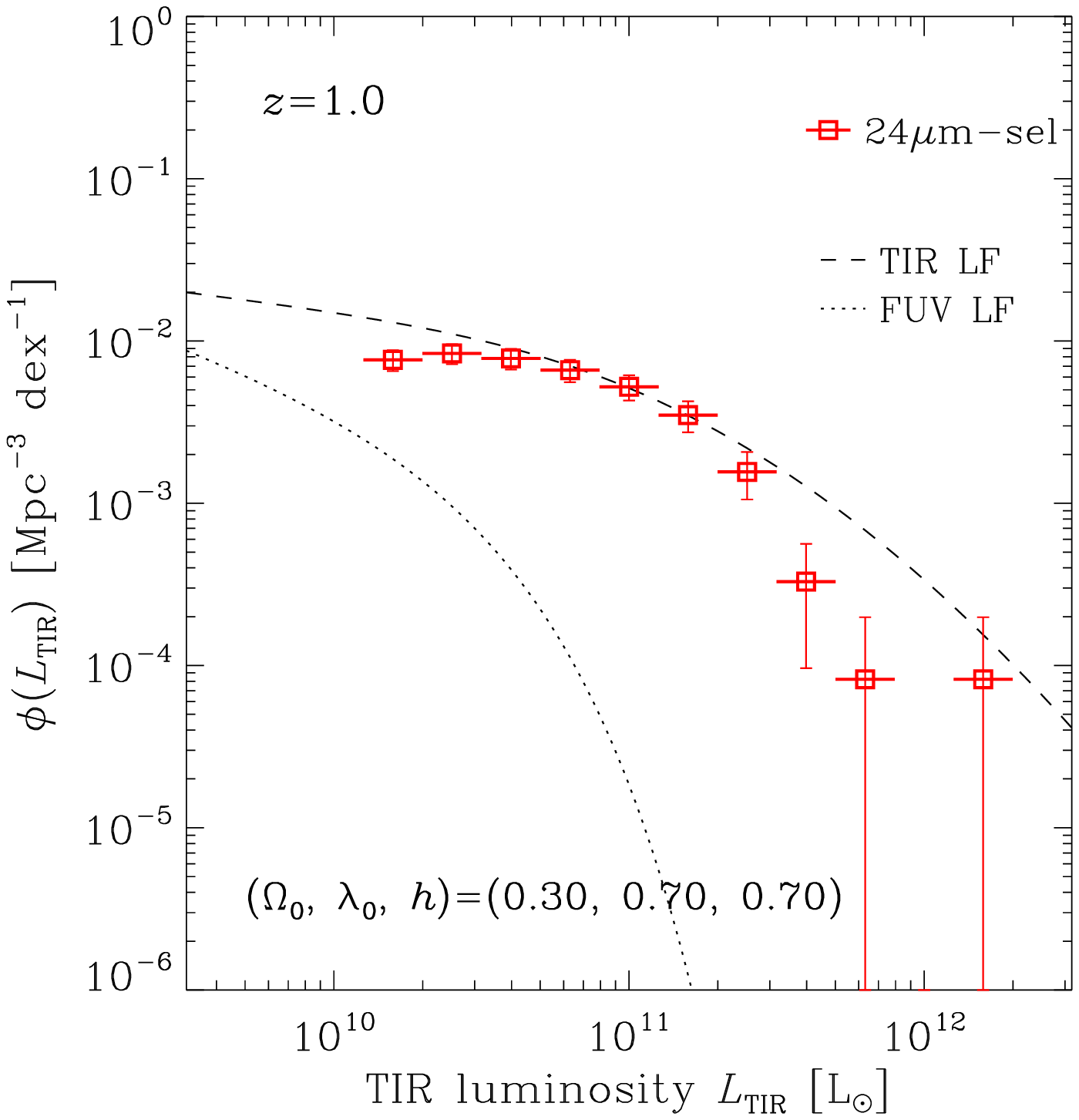}
\caption{The FUV and TIR LFs at $z = 1.0$.
Symbols are the same as in Fig.~\ref{fig:LF07}, but FUV samples are selected at
$U$-band.
}\label{fig:LF10} 
\end{figure*}

\section{Results and Discussion}
\label{sec:discussion}

\subsection{FUV and TIR univariate LFs}

In order to estimate the UV-IR BLF, first we have to examine our setting
for the FUV and TIR univariate LFs.
The {\sl IRAS}-{\sl GALEX} sample, the validity of the Local univariate LFs are already
proved [see Fig.~3 of Buat et al.\ (2007)].
Hence, we can safely use the analytic formulae of FUV and
TIR LFs at $z = 0$ [eqs.~(\ref{eq:schechter}) and (\ref{eq:saunders})]. 

We use the Schechter parameters presented by Wyder et al.\ (2005) for {\it GALEX} FUV: 
$(\alpha_1, L_{*1}, \phi_{*1}) = (1.21, 1.81\times 10^9h^{-2}\;{\rm [L_\odot]}, 
1.35 \times 10^{-2}h^3\;[\mbox{Mpc}^{-3}])$.
For the TIR, we used the parameters 
$(\alpha_2, L_{*2}, \phi_{*2}, \sigma) =  (1.23, 4.34 \times 10^8 h^{-2} \;{\rm [L_\odot]}, 
2.34 \times 10^{-2} h^3\;[\text{Mpc}^{-3}], 0.724)$ (Takeuchi et al.\ 2003) 
obtained from the {\sl IRAS} PSC$z$ galaxies (Saunders et al.\ 2000), and 
multiplied $L_{*1}$ with 2.5 to convert the 60-$\mu$m LF to the TIR LF.

For higher redshifts, ideally we should estimate the univariate FUV and TIR LFs 
simultaneously with the BLF estimation.
It is, however, quite difficult for our current samples because of the limited number of
galaxies.
We instead used the LF parameters at $z = 0.7$ and 1.0 obtained by 
previous studies on univariate LFs and modeled the FUV and TIR univariate LFs
at these redshifts and examined their validity with nonparametric LFs estimated from
the data.
We use the parameters compiled by Takeuchi et al.\ (2005a).
Parameters of the evolution of the TIR LF are obtained by approximating the
evolution in the form
\begin{eqnarray}
  \phi_2^{(1)}(L_2,z) = {\mathfrak g}(z) \phi_{2,0}^{(1)} \left[ \frac{L_2}{{\mathfrak f}(z)} \right]
\end{eqnarray}
where $\phi_{2,0}^{(1)} (z)$ is the local functional form of the TIR LF. 
Le Floc'h et al.\ (2005) assumed a power-law form for the evolution functions as
\begin{eqnarray}
   {\mathfrak f}(z) = (1+z)^{\mathfrak Q}, \; {\mathfrak g}(z) = (1+z)^{\mathfrak P} 
\end{eqnarray}
and obtained ${\mathfrak P} = 0.7$ and ${\mathfrak Q} = 3.2$, with $\alpha$ remaining constant.
The Schechter parameters at $z = 0.7$ and $1.0$ { for the FUV LF} are directly estimated 
by Arnouts et al.\ (2005) and we adopt their values (see Table 1 of Takeuchi et al.\ 2005a).

Then we estimated the FUV and TIR LFs with the stepwise maximum likelihood method and 
the variant of $C^-$ method from our CDFS multiwavelength data (for the estimation method, 
see, e.g., Takeuchi et al.\ 2000; Johnston 2011, and references therein).
The obtained univariate LFs at $z=0.7$ and 1.0 are presented in Figs.~\ref{fig:LF07} and \ref{fig:LF10}.
We also show the analytic model LFs in these figures.
We note that a well-known large density enhancement locates in the CDFS at $z=0.7$ (e.g., Salimbeni et al.\ 2009),
{ and thus} we have renormalized the LFs to remove the effect of the overdensity.
In the following analysis, we normalize the univariate LFs according to eq.~(\ref{eq:lfnorm}) so that
we can treat the univariate LFs as PDFs, hence this does not affect the following analysis at all.

Because of the small sample size, the LF shape does not perfectly agree with the supposed functional forms,
but the nonparametric LFs { are} acceptably similar to the analytic functions both for the FUV and TIR
at each redshift.
We stress that the analytic functions are {\it not} the fit to the data, but estimated from other
studies.
This implies that the estimated evolutionary parameters of the LFs work generally well.
Thus, we can use the higher-$z$ univariate LFs as the marginal PDFs for the
estimation of the UV-IR BLFs.

\begin{figure}[t]
\centering\includegraphics[width=0.45\textwidth]{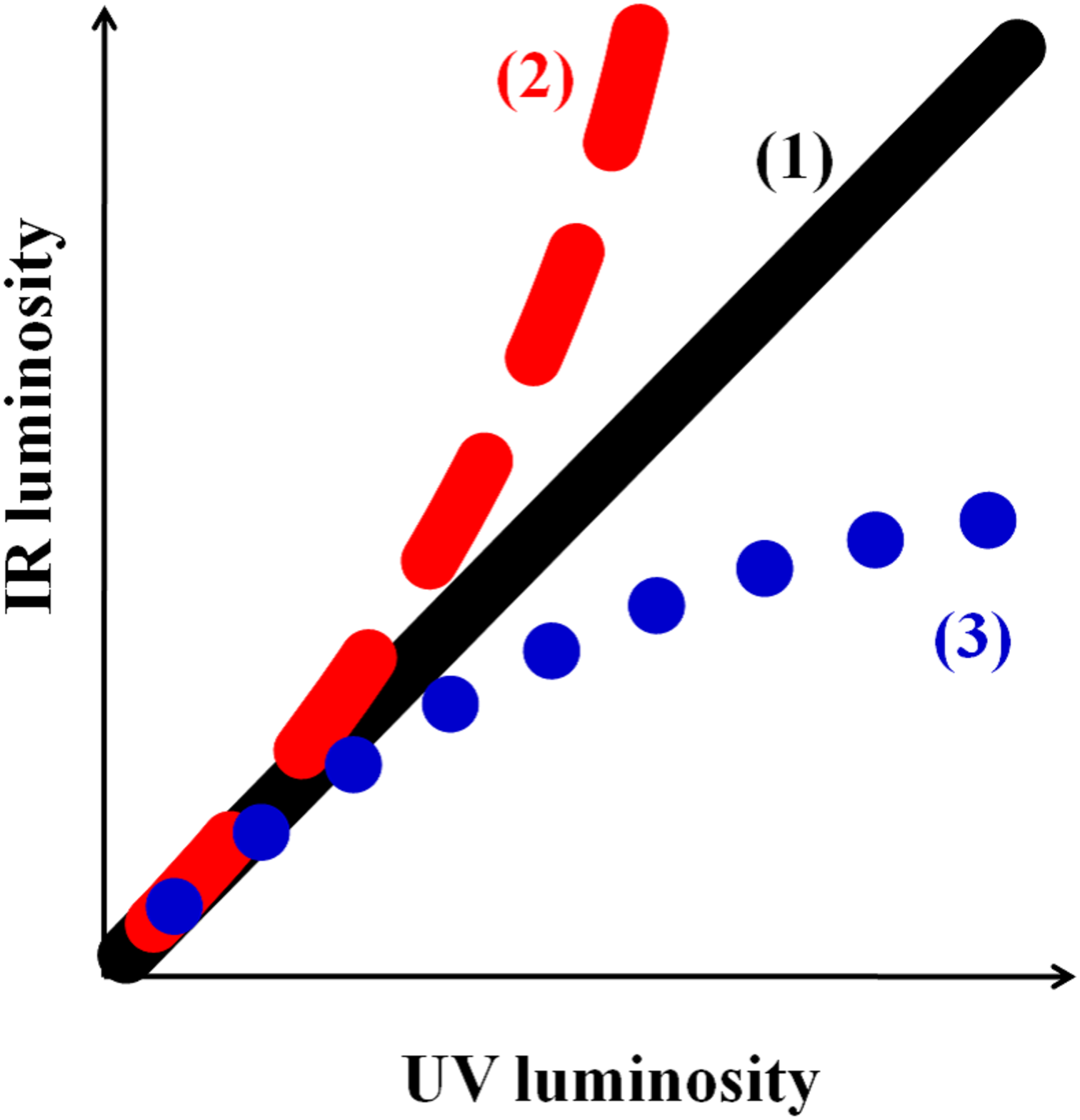}
\caption{Schematic BLF. 
(1) Diagonal: The energy from SF is emitted equally at UV and IR with any SF activity.
(2) Upward: The more active the SF  in a galaxy is, the more luminous at the IR (dusty SF).
(3) Downward: The more active the SF is, the more luminous at the UV (gtransparenth SF).
}\label{fig:schematicBLF}
\end{figure}

\subsection{Copula likelihood for the BLF estimation}
\label{subsec:likelihood}

By using the estimated univariate FUV and TIR LFs as given marginals, 
we can estimate only one parameter, the linear correlation $\rho$ by 
maximizing the likelihood function.
The structure of a { set of} two-band selected data is $(L_\fuv^{i_k}, j_\text{UV}^{i_k}, L_\tir^{i_k}, j_\text{IR}^{i_k}), i_k = 
1 \dots n_k$.
Here, $j_\text{band}$ (band: UV or IR) stands for the upper limit flag: 
$j_\text{band}=0$: detection and $j_\text{band}=-1$: upper limit.
Another index $k$ is the indicator of the selected band, i.e., $k=1$ means a sample 
galaxy is selected at UV and $k=-1$ means it is selected at IR.
The likelihood function is as follows.
\begin{eqnarray}\label{eq:likelihood}
  &&\ln {\cal L} \left(\left. L_{\fuv}^{i_k}, L_{\tir}^{i_k} \right| i_{k} = 1, \cdots n_{k}, k=1, -1 \right) \nonumber \\
    &&= \sum_{k=\left\{ {\tiny \begin{array}{rl} 1 & : {\rm UV \;sel} \\ -1 & : {\rm IR\;sel} \end{array}}\right. } 
       \sum_{i_k = 1}^{n_k} \Biggl\{ 
       \ln \left[ p^\text{det} \left( L_{\fuv}^{i_k}, L_{\tir}^{i_k} \right) \right]^{(1+j_\text{UV}^{i_k})(1+j_\text{IR}^{i_k})} \nonumber \\
    &&\quad+ \ln \left[ p^\text{UL: UV} \left( L_{\fuv}^{i_k}, L_{\tir}^{i_k} \right)\right]^{\frac{(1+k)(-j_\text{UV}^{i_k})}{2}} \nonumber \\
    &&\quad+ \ln \left[ p^\text{UL: IR} \left( L_{\fuv}^{i_k}, L_{\tir}^{i_k} \right)\right]^{\frac{(1-k)(-j_\text{IR}^{i_k})}{2}} \Biggr\},
\end{eqnarray}
where $p^\text{det}\left( L_{\fuv}^{i_k}, L_{\tir}^{i_k} \right) $ is the probability for $i_k$th galaxy to be 
detected at both bands and to have
luminosities $L_{\fuv}^{i_k}$ and $L_{\tir}^{i_k}$, 
\begin{eqnarray}\label{eq:survival}
  &&p^\text{det} \left( L_{\fuv}^{i_k}, L_{\tir}^{i_k} \right) \nonumber \\
    &&\equiv \dfrac{\phi^{(2)} \left( L_{\fuv}^{i_k}, L_{\tir}^{i_k} ; \rho \right)}{
    \displaystyle \int_{{L_\text{FUV}^\text{lim}}(z_{i_k})}^\infty \int_{L_\text{TIR}^{\text{lim}}(z_{i_k})}^\infty
    \phi^{(2)} \left( L_{\fuv}', L_{\tir}' ; \rho \right) 
    \pd L_{\tir}'\pd L_{\fuv}' }, \nonumber \\
\end{eqnarray}
$p^\text{UL: UV} \left( L_{\fuv}^{i_k}, L_{\tir}^{i_k} \right) $ is the probability for $i_k$th galaxy to 
be detected at IR band and have a luminosity $L_{\tir}^{i_k}$, but not detected at UV band and
only an upper limit $L_{\fuv, j_k}^{i_k}$ is available,
\begin{eqnarray}
  &&p^\text{UL: UV} \left( L_{\fuv}^{i_k}, L_{\tir}^{i_k} \right) \nonumber \\
    &&\equiv \dfrac{\displaystyle \int_{0}^{L_{\fuv,j_k}^{i_k}} \phi^{(2)} \left( L_\fuv', L_\tir^{i_k} \right) {\rm d}L_{\fuv}' }{
    \displaystyle \int_0^\infty \int_{L_\text{TIR}^{\text{lim}}(z_{i_k})}^\infty
    \phi^{(2)} \left( L_{\fuv}', L_{\tir}' ; \rho \right) 
    \pd L_{\tir}'\pd L_{\fuv}' },\nonumber \\
\end{eqnarray}
and $p^\text{UL: IR} \left( L_{\fuv}^{i_k}, L_{\tir}^{i_k} \right) $ is the probability for $i_k$th galaxy to 
be detected at UV band and have a luminosity $L_{\fuv}^{i_k}$, but not detected at IR band and
only an upper limit $L_{\tir, j_k}^{i_k}$ is available,
\begin{eqnarray}
  &&p^\text{UL: IR} \left( L_{\fuv}^{i_k}, L_{\tir}^{i_k} \right) \nonumber \\
    &&\equiv \dfrac{\displaystyle \int_{0}^{L_{\tir,j_k}^{i_k}} \phi^{(2)} \left( L_\fuv^{i_k}, L_\tir' \right) {\rm d}L_{\tir}' }{
    \displaystyle \int_{L_\text{FUV}^{\text{lim}}(z_{i_k})}^\infty \int_0^\infty 
    \phi^{(2)} \left( L_{\fuv}', L_{\tir}' ; \rho \right) 
    \pd L_{\tir}'\pd L_{\fuv}' }.\nonumber \\
\end{eqnarray}
The denominator in eq.~(\ref{eq:survival}) is introduced to take into account 
the truncation in the data by observational flux selection limits at both bands
(e.g., Sandage et al.\ 1979; Johnston 2011).
We should also note that it often happens that the same galaxies are 
included { in both the} UV- and IR-selected sample.
In such a case, we should count the same galaxies only once to avoid
double counting of them.
Practically, such galaxies are included in any of the samples, because they
are detected at both bands and are symmetric between UV- and IR-selections.

{ In the estimation procedure, we estimated the univariate LFs and their evolutions first, 
and then used these parameters to estimate the BLF and its evolution. 
One might wonder if this compromising method would introduce some bias in 
the estimation.
Since we fix the form of the maximum likelihood estimator 
eq.~(\ref{eq:likelihood}), this two-step estimation instead of simultaneous 
estimation does not bias the result, especially the dependence structure
between UV and IR, unless the assumed univariate LF shape would be
significantly different from the real one.
We have already seen that the stepwise maximum likelihood estimation gave
nonparametric LFs which reasonably agree with the assumed Schechter
or Saunders function.
Then, we can safely rely on the result for further discussion.

However, we should note that the two-step estimation gives conditional errors of the 
parameters only for each step, not the marginal ones. 
Then, the errors for each parameters are significantly underestimated.
If we want to discuss the error values more precisely, we need a larger dataset
and must use the simultaneous estimation method.
This will be done in the future works.
}

\subsection{The BLF and its evolution}

Using the Gaussian copula, now we can estimate the BLF.
The visible and hidden SFRs should be directly reflected to this function. 
Dust is produced by SF activity, but also destroyed by SN blast waves as a result of the SF. 
Many physical processes are related to the evolution of the dust amount. 
Thus, first of all, we should describe statistically how it evolved, as stated
in Introduction. 

First, we summarize how to interpret the UV-IR BLF schematically in Fig.~\ref{fig:schematicBLF}.
First, we see the case that the ridge of the BLF is straight and diagonal [see (1) in Fig.~\ref{fig:schematicBLF}].
This means that the energy from SF is emitted equally at UV and IR with any SF activity.
If the relation is diagonal but has an offset horizontally or vertically, this suggests that 
a constant fraction of energy is absorbed by dust and reemitted.
Second, if the ridge is curved upward, it means that the more active the SF in a galaxy is, 
the more luminous at the IR [dusty SF: (2) in Fig.~\ref{fig:schematicBLF}].
Third, if the ridge is bent downward, the more active the SF is, the more luminous at 
UV [`transparent' SF: (3) in Fig.~\ref{fig:schematicBLF}].

\begin{figure}[t]
\centering\includegraphics[width=0.45\textwidth]{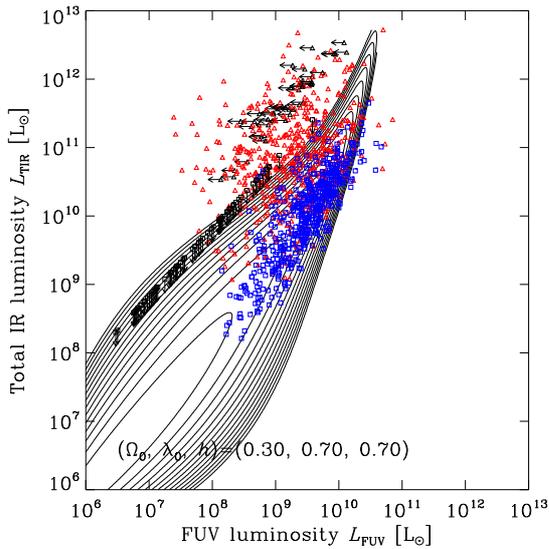}
\caption{The BLF of galaxies from {\sl IRAS} and {\sl GALEX} at $z = 0$.
Contours are the analytic model constructed by a Gaussian copula and
univariate FUV and TIR LFs.
{ Open squares represent the UV-selected sample from {\sl GALEX}, while
open triangles are the IR-selected ones from {\sl IRAS}.
Squares and triangles with arrows mean that they are upper limits at FUV and FIR, respectively. }
}\label{fig:blf_z0}
\end{figure}
\begin{figure}[t]
\centering\includegraphics[width=0.45\textwidth]{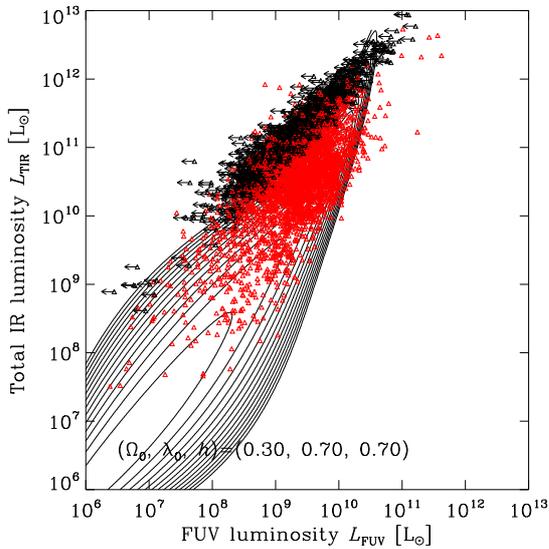}
\caption{The same as Fig.~\ref{fig:blf_z0} but from {\sl AKARI} and {\sl GALEX}.
{ Open triangles represent the IR-selected samples from {\sl AKARI}. 
Since there is no UV-selected sample in this figure, we only show the IR-selected
ones.
}
}\label{fig:blf_z0_akari}
\end{figure}
\begin{figure}[t]
\centering\includegraphics[width=0.45\textwidth]{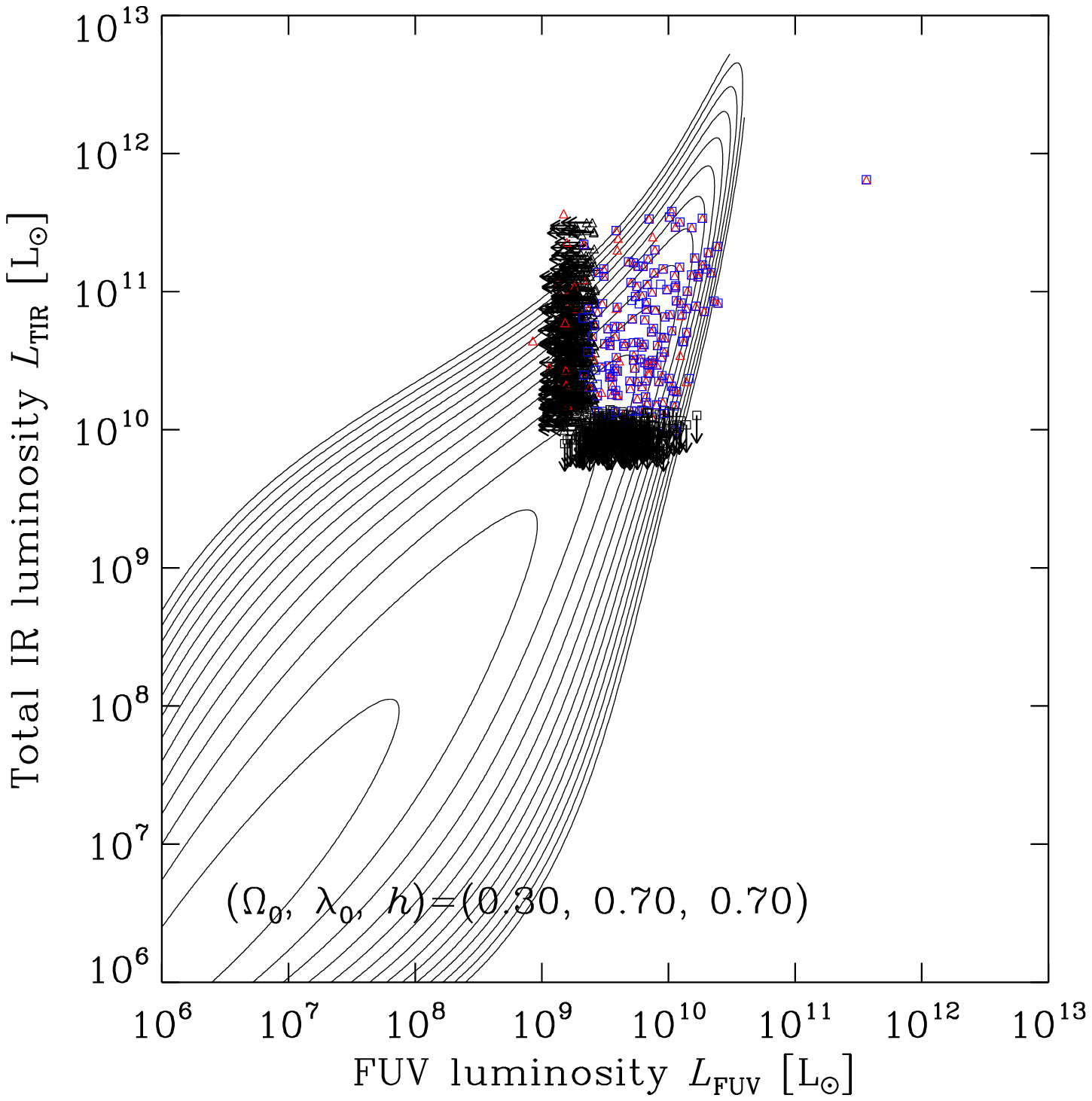}
\caption{The same as Fig.~\ref{fig:blf_z0} but from {\sl Spitzer} and {\sl GALEX} at $z = 0.7$.
{ Symbols are essentially the same as in Fig.~\ref{fig:blf_z0}, but IR-selected samples
are from {\sl Spitzer}/MIPS.
}
}\label{fig:blf_z07}
\end{figure}
\begin{figure}[t]
\centering\includegraphics[width=0.45\textwidth]{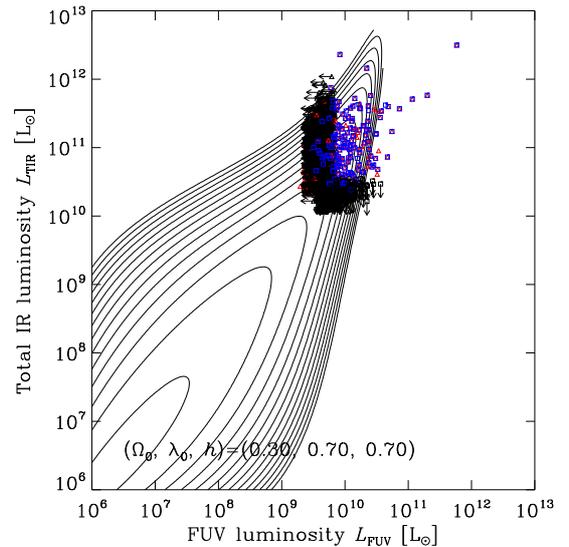}
\caption{The same as Fig.~\ref{fig:blf_z0} but from {\sl Spitzer} and {\sl GALEX} at $z = 1.0$.
{ Again symbols are essentially the same as in Fig.~\ref{fig:blf_z0}, but IR-selected samples
are from {\sl Spitzer}/MIPS.
}
}\label{fig:blf_z10}
\end{figure}

Now we show the estimated BLFs in Figs.~\ref{fig:blf_z0}--\ref{fig:blf_z10}.
In the Local Universe, the BLF is quite well constrained.
The estimated correlation coefficient $\rho$ is very high: 
$\rho = 0.95 \pm 0.04$ for {\sl IRAS}-{\sl GALEX}
and $\rho = 0.95 \pm 0.006$ for {\sl AKARI}-{\sl GALEX} datasets.
The apparent scatter of the $L_{\rm FUV}$--$L_{\rm TIR}$ is 
found to be due to the nonlinear shape of the ridge of the BLF. 
This bent shape of the BLF was implied by preceding studies (Martin et al.\ 2005),
and we could quantify this feature. 
The copula BLF naturally reproduced it. 
{ For the {\sl AKARI}-{\sl GALEX} sample, only the IR-selected sample
is available at this moment.
This, however, does not bias the BLF estimation since the information from censored
data points are incorporated properly in eq.~(\ref{eq:likelihood}). 
This can be directly tested, for example, by using only one of the UV and IR-selected data for
the $z=0$ BLF estimation with {\sl IRAS}-{\sl GALEX} dataset.
Both one-band estimations with UV and IR-selected data yield $\rho = 0.95 \pm 0.07$.
Obviously the error becomes larger because of fewer data, but the estimate itself remain
unchanged.
}

At higher redshifts ($z = 0.7\mbox{--}1.0$), the linear correlation remains tight 
($\rho \simeq 0.91 \pm 0.06$ at $z = 0.7$ and $\rho \simeq 0.86 \pm 0.05$ at $z = 1.0$), 
even though it is difficult to constrain the low-luminosity end from the data 
in this analysis ({\sl Spitzer}-{\sl GALEX} in the CDFS). 
{ The distribution of upper limits in Figs.~\ref{fig:blf_z07} and \ref{fig:blf_z10}
looks different from that in Fig.~\ref{fig:blf_z0}.
Since we have restricted the redshifts of galaxies in these samples, the redshift
restriction gives approximately constant luminosity limits both at FUV and FIR.
This gives the ``L-shaped'' distribution of upper limits seen in these figures.
}

Though the whole shape cannot be perfectly determined by the current data,
we find that $\rho$ in the copula LF is high and remarkably stable with redshifts
(from 0.95 at $z = 0$ to 0.85 at $z = 1.0$).
This implies the evolution of the UV-IR bivariate LF is mainly due to 
the different evolution of the univariate LFs, and may not be controled 
by the evolution of the dependence structure.

{ Forthcoming better data in the future will improve various aspects of the
BLF estimation.
First, {\sl Herschel}, ALMA, and 
{\sl SPICA}\footnote{$^7$ URL: http://www.ir.isas.jaxa.jp/SPICA/SPICA{\_}HP/index{\_}English.html.} 
will provide us with direct estimations of $L_{\rm TIR}$ of galaxies especially at high redshifts,
since they will detect these galaxies at longer wavelengths than 24~$\mu$m.
This allows us to avoid a possible bias of the estimated $L_{\rm TIR}$ caused by the 24~$\mu$m-based 
extrapolation of the SED. 

Second, when we have deeper data at UV and IR, we will be able to extend the luminosity range 
toward much lower luminosities both at FUV and FIR. 
Then, the faint-end structure of the BLF will be much more tightly determined.
Further, it is not only for the determination of the faint end, but deeper data will 
give some insights for the choice of the copula functional form.
In this study, we did not try to examine if the Gaussian copula would be 
a proper choice as a model of the UV-IR BLF.
At $z = 0$, when UV-selected data comparably large as 
the {\sl AKARI} sample are ready, we will be able to 
specify (a family of) copulas appropriate for this analysis.
At higher redshifts, our current data are not deep enough to have 
any claim on the faint-end dependence structure of the BLF, but 
again {\sl Herschel}, ALMA, and {\sl SPICA} data together with ground-based
optical or {\sl JWST} ones will enables us to examine which copula would be
appropriate to describe the BLF, and constrain the evolution of the BLF along
the whole history of galaxy evolution in the Universe.
}

\section{Conclusion}\label{sec:conclusion}

To understand the visible and hidden SF history in the Universe, 
it is crucial to analyze multiwavelength data in a unified manner. 
The copula is an ideal tool to combine two marginal univariate 
LFs to construct a bivariate LFs.
It is straightforward to extend this method to multivariate DFs.
\begin{enumerate}{ 
\item The Gaussian copula LF is sensitive to the linear correlation parameter $\rho$. 
\item Even so, $\rho$ in the copula LF is remarkably stable with redshifts (from 0.95 at $z = 0$ to 0.85 at $z = 1.0$).
\item This implies the evolution of the UV-IR BLF is mainly due to 
the different evolution of the univariate LFs, and may not be controled by the dependence structure.
\item The nonlinear structure of the BLF is naturally reproduced by the Gaussian copula.}
\end{enumerate}
The data from {\sl Herschel}, ALMA, and {\sl SPICA} data will improve the estimates drastically,
and we expect to specify the full evolution of the UV-IR BLF in the Universe.
We stress that the copula will be a useful tool for any other kind of 
bi- (multi-) variate statistical analysis.

\acknowledgments
This work is based on observations with {\sl AKARI}, a JAXA project with the participation of ESA. 
This research has made use of the NASA/IPAC Extragalactic Database (NED) which is operated 
by the Jet Propulsion Laboratory, California Institute of Technology, under contract with 
the National Aeronautics and Space Administration. 
TTT has been supported by Program for Improvement of Research
Environment for Young Researchers from Special Coordination Funds for
Promoting Science and Technology, and the Grant-in-Aid for the Scientific
Research Fund (20740105, 23340046, 24111707) commissioned by the MEXT. 
TTT, AS, and FTY have been partially supported from the Grand-in-Aid for the Global
COE Program ``Quest for Fundamental Principles in the Universe: from
Particles to the Solar System and the Cosmos'' from the Ministry of
 Education, Culture, Sports, Science and Technology (MEXT) of Japan.


\email{T. T. Takeuchi (e-mail: takeuchi.tsutomu@g.mbox.nagoya-u.ac.jp)}
\label{finalpage}
\lastpagesettings

\begin{references}\frenchspacing
\bibitem[Arnouts et al.(2001)]{arnouts01}
 Arnouts, S., et al., ESO imaging survey. Deep public survey: Multi-color optical data for 
 the Chandra Deep Field South, \textit{Astronomy and Astrophysics}, 
 \textbf{379}, 740--754, 2001. 
\bibitem[Arnouts et al.(2005)]{arnouts05} 
 Arnouts, S., et al., The {\sl GALEX} VIMOS-VLT Deep Survey Measurement of the 
 Evolution of the 1500 {\AA} Luminosity Function, \textit{The Astrophysical Journal},
 \textbf{619}, L43--L46, 2005. 
\bibitem[Asano et al.(2012)]{asano11}
 Asano, R. S., Takeuchi, T.~T., Hirashita, H., Inoue, A. K., Dust formation history of galaxies: a critical role of
 metallicity for the dust mass growth by accreting materials in the interstellar
 medium, \textit{Earth, Planets, \& Space}, 2012, in press (astro-ph/1206.0817)
\bibitem[Benabed et al.(2009)]{benabed09} 
 Benabed, K., Cardoso, J.-F., Prunet, S., Hivon, E., TEASING: a fast and accurate 
 approximation for the low multipole likelihood of the cosmic microwave 
 background temperature, \textit{Monthly Notices of the Royal Astronomical Society}, 
 \textbf{400}, 219--227, 2009.
\bibitem[Bothwell et al.(2011)]{bothwell11} 
 Bothwell, M.~S., Kennicutt, R.~C., Johnson, B.~D., Wu, Y., Lee, J.~C., Dale, D., Engelbracht, 
 C., Calzetti, D., Skillman, E., The star formation rate distribution 
 function of the local Universe, \textit{Monthly Notices of the Royal Astronomical 
 Society}, \textbf{415}, 1815--1826, 2011. 
\bibitem[Buat \& Burgarella(1998)]{buat98} 
 Buat, V., Burgarella, D., The observation of the nearby universe in UV and in FIR: 
 an evidence for a moderate extinction in present day star forming galaxies, \textit{Astronomy and Astrophysics}, 
 \textbf{334}, 772--782, 1998.
\bibitem[Buat et al.(2005)]{buat05}
 Buat, V., et al. Dust Attenuation in the Nearby Universe: A Comparison 
 between Galaxies Selected in the Ultraviolet and in the Far-Infrared, 
 \textit{The Astrophysical Journal}, \textbf{619}, L51--L54, 2005. 
\bibitem[Buat et al.(2007)]{buat07} 
 Buat, V., Takeuchi, T.~T., Iglesias-P{\'a}ramo, J., et al., The Local Universe as Seen 
 in the Far-Infrared and Far-Ultraviolet: A Global Point of View of the Local Recent Star Formation, 
 \textit{The Astrophysical Journal Supplement Series}, \textbf{469}, 19--25, 2007. 
\bibitem[Buat et al.(2009)]{buat09} 
 Buat, V., Takeuchi, T.~T., Burgarella, D., Giovannoli, E., Murata, K.~L., 
 The infrared emission of ultraviolet-selected galaxies from $z = 0$ to $z = 1$,
 \textit{Astronomy and Astrophysics}, \textbf{507}, 693--704. 2009. 
\bibitem[Burgarella et al.(2006)]{burgarella06}
 Burgarella, D., et al., Ultraviolet-to-far infrared properties of Lyman break 
 galaxies and luminous infrared galaxies at $z \sim 1$, \textit{Astronomy and Astrophysics}, 
 \textbf{450}, 69--76, 2006. 
\bibitem[Cortese et al.(2006)]{cortese06} 
 Cortese, L., Boselli, A., Buat, V., Gavazzi, G., Boissier, S., Gil de Paz, A., Seibert, M.,  
 Madore, B.~F., Martin, D.~C.\ 2006.\ UV Dust Attenuation in Normal 
 Star-Forming Galaxies. I. Estimating the $L_{\rm TIR}/L_{\rm FUV}$ Ratio, \textit{The 
 Astrophysical Journal}, \textbf{637}, 242--254, 2006.
\bibitem[Cucciati et al.(2012)]{cucciati12}
 Cucciati, O., et al.\ The star formation rate density and dust attenuation evolution over 12 Gyr
 with the VVDS surveys, \textit{Astronomy and Astrophysics}, \textbf{539}, A31, 2012. 
\bibitem[Dale et al.(2001)]{dale01} 
 Dale, D.~A., Helou, G., Contursi, A., Silbermann, N.~A., Kolhatkar, S., The Infrared 
 Spectral Energy Distribution of Normal Star-forming Galaxies, \textit{The 
 Astrophysical Journal}, \textbf{549}, 215--227, 2001. 
\bibitem[Driver et al.(2011)]{driver11}
 Driver, S.~P., et al., Galaxy and Mass Assembly (GAMA): survey diagnostics
 and core data release, \textit{Monthly Notices of the Royal Astronomical Society},
 \textbf{413}, 971--995, 2011. 
\bibitem[Dwek and Scalo(1980)]{dwek80} 
 Dwek, E., Scalo, J.~M., The evolution of refractory interstellar grains in the solar 
 neighborhood, \textit{The Astrophysical Journal}, \textbf{239}, 193--211, 1980. 
\bibitem[Dwek(1998)]{dwek98} 
 Dwek, E., The Evolution of the Elemental Abundances in the Gas and Dust 
 Phases of the Galaxy, \textit{The Astrophysical Journal}, \textbf{501}, 643, 1998. 
\bibitem[Eales et al.(2010)]{eales10}
 Eales, S., et al. The {\sl Herschel} ATLAS, \textit{Publications of the Astronomical 
 Society of the Pacific}, \textbf{122}, 499--515, 2010.
\bibitem[Elbaz et al.(2007)]{elbaz07}
 Elbaz, D., et al., The reversal of the star formation-density relation in the 
 distant universe, \textit{Astronomy and Astrophysics}, \textbf{468}, 33--48, 2007. 
\bibitem[Haines et al.(2011)]{haines11} 
 Haines, C.~P., Busarello, G., Merluzzi, P., Smith, R.~J., Raychaudhury, S., Mercurio, A., 
 Smith, G.~P ACCESS - II. A complete census of star formation in 
 the Shapley supercluster - UV and IR luminosity functions, \textit{Monthly Notices 
 of the Royal Astronomical Society}, \textbf{412}, 127-144, 2011. 
\bibitem[Hao et al.(2011)]{hao11} 
 Hao, C.-N., Kennicutt, R.~C., Johnson, B.~D., Calzetti, D., Dale, D.~A., Moustakas, J.,
 Dust-corrected Star Formation Rates of Galaxies. II. Combinations of 
 Ultraviolet and Infrared Tracers, \textit{The Astrophysical Journal}, \textbf{741}, 124 (22pp), 2011. 
\bibitem[Iglesias-P{\'a}ramo et al.(2006)]{iglesias06} 
 Iglesias-P{\'a}ramo, J., et al., Star Formation in the Nearby Universe:
 The Ultraviolet and Infrared Points of View, \textit{The Astrophysical Journal Supplement Series},
 \textbf{164}, 38--51, 2006. 
\bibitem[Johnston(2011)]{johnston11} 
 Johnston, R., Shedding light on the galaxy luminosity function, \textit{Astronomy and Astrophysics Review},
 \textbf{19}, 41, 2011.
\bibitem[Kawada et al.(2007)]{kawada07} 
 Kawada, M., et al., The Far-Infrared Surveyor (FIS) for {\sl AKARI},
 \textit{Publications of the Astronomical Society of Japan}, \textbf{59}, 389--400, 2007.
\bibitem[Koen(2009)]{koen09} 
 Koen, C., Confidence intervals for the correlation between the gamma-ray burst peak energy and 
 the associated supernova peak brightness, \textit{Monthly Notices of the Royal 
 Astronomical Society}, \textbf{393}, 1370-1376, 2009.
\bibitem[Le Floc'h et al.(2005)]{lefloch05}
 Le Floc'h, E., et al., Infrared Luminosity Functions from the Chandra
 Deep Field-South: The {\sl Spitzer} View on the History of Dusty Star Formation at
 $0 \leq z \leq 1$, \textit{The Astrophysical Journal}, \textbf{632}, 169--190, 2005. 
\bibitem[Martin et al.(2005)]{martin05} 
 Martin, D.~C., et al. The Star Formation Rate Function of the Local Universe,  
 \textit{The Astrophysical Journal}, \textbf{619}, L59--L62, 2005. 
\bibitem[Morrissey et al.(2007)]{Morrissey07} 
 Morrissey, P., et al., The Calibration and Data Products of {\sl GALEX},
 \textit{Astronomy and Astrophysics}, \textbf{173}, 682--697, 2007.
\bibitem[Murphy et al.(2011)]{murphy11} 
 Murphy, E.~J., Chary, R.-R., Dickinson, M., Pope, A., Frayer, D.~T., An 
 Accounting of the Dust-obscured Star Formation and Accretion Histories Over 
 the Last $\sim 11$ Billion Years, \textit{The Astrophysical Journal}, \textbf{732}, 126 (17pp), 2011. 
\bibitem[Nozawa et al.(2003)]{nozawa03} 
 Nozawa, T., Kozasa, T., Umeda, H., Maeda, K., Nomoto, K., Dust in the Early Universe: Dust 
 Formation in the Ejecta of Population III Supernovae, \textit{The Astrophysical 
 Journal}, \textbf{598}, 785--803, 2003.
\bibitem[Paturel et al.(2003)]{paturel03}
 Paturel, G., Petit, C., Prugniel, P., Theureau, G., Rousseau, J., Brouty, M., Dubois, P., 
Cambr{\'e}sy, L., HYPERLEDA.  I. Identification and designation of galaxies,
\textit{Astronomy and Astrophysics}, \textbf{412}, 45--55, 2003. 
\bibitem[Sakurai et al.(2012)]{sakurai11}
 Sakurai, A., Takeuchi, T. T., Yuan, F.-T., Buat, V., Burgarella, D., 
 Star Formation and Dust Extinction Properties of Local Galaxies Seen from {\sl AKARI} and {\sl GALEX}, 
 \textit{Earth, Planets and Space}, submitted, 2011.
\bibitem[Salimbeni et al.(2009)]{salimbeni09}
 Salimbeni, S., et al. A comprehensive study of large-scale structures in the GOODS-SOUTH field up to 
 $z \sim 2.5$, \textit{Astronomy and Astrophysics}, \textbf{501}, 865--877, 2009. 
\bibitem[Sandage et al.(1979)]{sandage79} 
 Sandage, A., Tammann, G.~A., Yahil, A., The velocity field of bright nearby galaxies. I - 
 The variation of mean absolute magnitude with redshift for galaxies in a 
 magnitude-limited sample, \textit{The Astrophysical Journal},
 \textbf{232}, 352--364, 1979. 
\bibitem[Sato et al.(2010)]{sato10} 
 Sato, M., Ichiki, K., Takeuchi, T.~T., Precise Estimation of Cosmological Parameters Using 
 a More Accurate Likelihood Function, \textit{Physical Review Letters}, \textbf{105}, 251301, 2010. 
\bibitem[Sato et al.(2011)]{sato11} 
 Sato, M., Ichiki, K., Takeuchi, T.~T., Copula cosmology: Constructing a likelihood 
 function, \textit{Physical Review D}, \textbf{83}, 023501, 2011. 
\bibitem[Saunders et al.(1990)]{saunders90} 
 Saunders, W., Rowan-Robinson, M., Lawrence, A., Efstathiou, G., Kaiser, N., Ellis, R.~S., 
 Frenk, C.~S., The 60-micron and far-infrared luminosity functions of 
 {\sl IRAS} galaxies, \textit{Monthly Notices of the Royal Astronomical Society}, \textbf{242}, 
 318--337, 1990.
\bibitem[Saunders et al.(2000)]{saunders00} 
 Saunders, W., et al., The PSCz catalogue, \textit{Monthly Notices of the Royal Astronomical Society}, \textbf{317},
 55--63, 2000. 
\bibitem[Schechter(1976)]{schechter76} 
Schechter, P., An analytic expression for the luminosity function for galaxies, \textit{The 
Astrophysical Journal}, \textbf{203}, 297--306, 1976.  
\bibitem[Scherrer et al.(2010)]{scherrer10} 
 Scherrer, R.~J., Berlind, A.~A., Mao, Q., McBride, C.~K., From Finance to Cosmology: 
 The Copula of Large-Scale Structure, \textit{The Astrophysical Journal}, \textbf{708}, L9-L13, 2010. 
\bibitem[Seibert et al.(2005)]{seibert05} 
 Seibert, M., et al., Testing the Empirical Relation between Ultraviolet Color 
 and Attenuation of Galaxies, \textit{The Astrophysical Journal}, \textbf{619}, L55--L58, 2005
\bibitem[Takeuchi et al.(2000)]{takeuchi00} 
 Takeuchi, T.~T., Yoshikawa, K., Ishii, T.~T., Tests of Statistical Methods for 
 Estimating Galaxy Luminosity Function and Applications to the Hubble Deep 
 Field, \textit{The Astrophysical Journal Supplement Series}, \textbf{129}, 1--31, 2000.
\bibitem[Takeuchi et al.(2001a)]{takeuchi01a} 
 Takeuchi, T.~T., Ishii, T.~T., Hirashita, H., Yoshikawa, K., Matsuhara, H., Kawara, K., 
 Okuda, H., Exploring Galaxy Evolution from Infrared Number Counts 
 and Cosmic Infrared Background, \textit{Publications of the Astronomical Society 
 of Japan}, \textbf{53}, 37--52, 2001a.
\bibitem[Takeuchi et al.(2001b)]{takeuchi01b} 
 Takeuchi, T.~T., Kawabe, R., Kohno, K., Nakanishi, K., Ishii, T.~T., Hirashita, H., 
 Yoshikawa, K., Impact of Future Submillimeter and Millimeter Large 
 Facilities on the Studies of Galaxy Formation and Evolution, \textit{Publications 
 of the Astronomical Society of the Pacific}, \textbf{113}, 586--606, 2001b.
\bibitem[Takeuchi et al.(2003)]{takeuchi03} 
 Takeuchi, T.~T., Yoshikawa, K., Ishii, T.~T.\ The Luminosity Function of {\sl IRAS} Point 
 Source Catalog Redshift Survey Galaxies, \textit{The Astrophysical Journal}, \textbf{587}, 
 L89-L92, 2003.
\bibitem[Takeuchi et al.(2005a)]{takeuchi05a}
 Takeuchi, T.~T., Buat, V., Burgarella, D.\ The evolution of the ultraviolet and 
 infrared luminosity densities in the universe at $0 < z < 1$,
 \textit{Astronomy and Astrophysics}, \textbf{440}, L17-L20, 2005a.
\bibitem[Takeuchi et al.(2005b)]{takeuchi05b} 
 Takeuchi, T.~T., Buat, V., Iglesias-P{\'a}ramo, J., Boselli, A., \&
 Burgarella, D., Mid-infrared luminosity as an indicator of the total
 infrared luminosity of galaxies, \textit{Astronomy and Astrophysics}, \textbf{432}, 423--429,
 2005b. 
\bibitem[Takeuchi et al.(2005c)]{takeuchi05c}
 Takeuchi, T.~T., Ishii, T.~T., Nozawa, T., Kozasa, T., Hirashita, H., A model for the 
 infrared dust emission from forming galaxies, \textit{Monthly Notices of the Royal 
 Astronomical Society}, \textbf{362}, 592-608, 2005c.
\bibitem[Takeuchi et al.(2006)]{takeuchi06} 
 Takeuchi, T.~T., Ishii, T.~T., Dole, H., Dennefeld, M., Lagache, G., 
 Puget, J.-L., The ISO 170 {$\mu$}m luminosity function of galaxies, 
 \textit{Astronomy and Astrophysics}, \textbf{448}, 525--534, 2006. 
\bibitem[Takeuchi(2010)]{takeuchi10a} 
 Takeuchi, T.~T., Constructing a bivariate distribution function with given marginals and 
 correlation: application to the galaxy luminosity function, \textit{Monthly 
 Notices of the Royal Astronomical Society}, \textbf{406}, 1830-1840, 2010.
\bibitem[Takeuchi et al.(2010)]{takeuchi10b} 
 Takeuchi, T.~T., Buat, V., Heinis, S., Giovannoli, E., Yuan, F.-T., 
 Iglesias-P{\'a}ramo, J., Murata, K.~L., Burgarella, D., 
 Star formation and dust extinction properties of local galaxies from the 
 {\sl AKARI}-{\sl GALEX} all-sky surveys. First results from the most secure multiband 
 sample from the far-ultraviolet to the far-infrared, \textit{Astronomy and Astrophysics},
 \textbf{514}, A4, 2010. 
\bibitem[Wang and Rowan-Robinson(2009)]{wang09} 
 Wang, L., Rowan-Robinson, M., The Imperial {\sl IRAS}-FSC Redshift Catalogue,
 \textit{Monthly Notices of the Royal Astronomical Society}, \textbf{398}, 109-118,
 2009. 
\bibitem[Wolf et al.(2004)]{wolf04}
 Wolf, C., et al., A catalogue of the Chandra Deep Field South with multi-colour
 classification and photometric redshifts from COMBO-17, \textit{Astronomy and Astrophysics},
 \textbf{421}, 913--936, 2004.
\bibitem[Wyder et al.(2005)]{wyder05}
 Wyder, T.~K., et al., The Ultraviolet Galaxy Luminosity Function in the Local 
 Universe from {\sl GALEX} Data, \textit{The Astrophysical Journal}, \textbf{619}, L15--L18, 
 2005. 
\bibitem[Yamamura et al.(2010)]{yamamura10}
 Yamamura I., Makiuti S., Ikeda N., Fukuda Y., Oyabu S., Koga T., White 
 G.~J., {\sl AKARI}/FIS All-Sky Survey Bright Source Catalogue Version 1.0
 Release Note, ISAS/JAXA, 2010.
\end{references}
\end{document}